\documentclass[aps,pra,floatfix,superscriptaddress,twocolumn]{revtex4}
\usepackage[dvips]{graphicx}

\usepackage{longtable}
\usepackage{dcolumn}
\usepackage[dvips]{graphicx}
\usepackage{bm}
\usepackage{bbm}

\usepackage{times}
\usepackage{nicefrac}
\usepackage{amsmath}
\usepackage{amsfonts}
\usepackage{amssymb}
\usepackage{amsthm}

\newcolumntype{.}{D{x}{}{-1}}

\newcommand{\balpha}{\bm{\alpha}}

\newcommand{\vare}{\varepsilon}
\newcommand{\bfr}{{\bm {r}}}

\newcommand{\lbr}{\langle}
\newcommand{\rbr}{\rangle}

\newcommand{\Za}{Z\alpha}

\begin{document}

\title{One-loop electron self-energy for the bound-electron $\bm{g}$ factor}

\author{V.~A. Yerokhin}
\affiliation{Max~Planck~Institute for Nuclear Physics, Saupfercheckweg~1, D~69117 Heidelberg,
Germany} \affiliation{Center for Advanced Studies, Peter the Great St.~Petersburg Polytechnic
University, 195251 St.~Petersburg, Russia}

\author{Z. Harman}
\affiliation{Max~Planck~Institute for Nuclear Physics, Saupfercheckweg~1, D~69117 Heidelberg,
Germany}

\begin{abstract}

We report calculations of the one-loop self-energy correction to the bound-electron $g$ factor of
the $1s$ and $2s$ states of light hydrogen-like ions with the nuclear charge number $Z \le 20$.
The calculation is carried out to all orders in the binding nuclear strength. We find good
agreement with previous calculations and improve their accuracy by about two orders of magnitude.

\end{abstract}

\pacs{31.30.jn, 31.15.ac, 32.10.Dk, 21.10.Ky}

\maketitle

The bound-electron $g$ factor in light hydrogen-like and lithium-like ions has been measured with a
high accuracy, which reached to $3\times 10^{-11}$ in the case of C$^{5+}$ \cite{sturm:14}. Such
measurements have yielded one of the best tests of the bound-state QED theory \cite{sturm:11} and
significantly improved the precision of the electron mass \cite{sturm:14,mohr:16:codata}. Further
advance of the experimental accuracy toward the $10^{-12}$ level is anticipated in the near future
\cite{sturm:17}.

One of the dominant effects in the bound-electron $g$ factor is the one-loop electron self-energy.
Its contribution to the total $g$ factor value is so large that the effect needs to be calculated
to all orders in the nuclear binding strength parameter $\Za$ even for ions as light as carbon ($Z$
is the nuclear charge number, $\alpha$ is the fine-structure constant). The numerical error in the
evaluation of the electron self-energy is currently the second-largest source of uncertainty for
the hydrogen-like ions (the largest error stemming from the two-loop electron self-energy
\cite{pachucki:04:prl,pachucki:05:gfact}). The error needs to be descreased in order to match the
anticipated experimental precision.

The numerical accuracy of the one-loop self-energy is also relevant for the determination of the
electron mass \cite{sturm:14,mohr:16:codata}. The self-energy values actually used in the
electron-mass determinations were obtained by an extrapolation of the high- and medium-$Z$
numerical results down to $Z = 6$ (carbon) and $8$ (oxygen). Clearly, this situation is not fully
satisfactory and a direct numerical calculation would be preferable.

All-order (in $\Za$) calculations of the electron self-energy to the bound-electron $g$ factor have
a long history. First calculations of this correction were accomplished two decades ago
\cite{persson:97:g,blundell:97,beier:00:pra}. The numerical accuracy of these evaluations was
advanced in the later works \cite{yerokhin:02:prl,yerokhin:04}, which was crucial at the time as it
brought an improvement of the electron mass determination. This correction was revisited again in
Refs.~\cite{yerokhin:08:prl,yerokhin:10:sehfs}. In the present work, we aim to advance the
numerical accuracy of the one-loop electron self-energy and bring it to the level required for
future experiments.

We consider the one-loop self-energy correction to the $g$ factor of an electron bound by the
Coulomb field of the point-like and spinless nucleus. This correction can be represented
\cite{yerokhin:02:prl,yerokhin:04} as a sum of the irreducible (ir) and the vertex$+$reducible (vr)
parts,
\begin{align}
\Delta g_{\rm SE} = \Delta g_{\rm ir} + \Delta g_{\rm vr}\,.
\end{align}
The irreducible part is
\begin{align}
\Delta g_{\rm ir} = 2\, \lbr \delta_g a| \gamma^0 \widetilde{\Sigma}(\vare_a)|a\rbr\,,
\end{align}
where $\widetilde{\Sigma}(\vare) = \Sigma(\vare)-\delta m$ is the (renormalized) one-loop
self-energy operator (see, e.g., \cite{yerokhin:10:sehfs}) and $| \delta_g a\rbr$ is the perturbed
wave function
\begin{align}
| \delta_g a\rbr = \sum_{n \neq a} \frac{|n\rbr\lbr n|\delta V_g|a\rbr}{\vare_a-\vare_n}\,,
\end{align}
with $\delta V_g = 2m[\bfr \times \balpha]_z$ being the effective $g$-factor operator
\cite{yerokhin:10:sehfs} that assumes that the spin projection of the reference state is $m_a=
1/2$. The vertex$+$reducible part is
\begin{align}\label{eq1}
\Delta g_{\rm vr} = &\ \frac{i}{2\pi} \int_{C}d\omega\, \sum_{n_1n_2} \biggl[
  \frac{\lbr n_1|\delta V_g |n_2\rbr\lbr an_2|I(\omega)|n_1a\rbr}{(\Delta_{an_1}-\omega)(\Delta_{an_2}-\omega)}
\nonumber \\ &
   -
  \delta_{n_1n_2}\,\frac{\lbr a|\delta V_g |a\rbr\lbr an_1|I(\omega)|n_1a\rbr}{(\Delta_{an_1}-\omega)^2}
  \biggr]\,,
\end{align}
where  $I(\omega)$ is the operator of the electron-electron interaction (see, e.g.,
\cite{yerokhin:04}), $\omega$ is the energy of the virtual photon, $\Delta_{ab} = \vare_a-\vare_b$,
and a proper covariant identification and cancellation of ultraviolet and infrared divergences is
assumed. The integration contour $C$ in Eq.~(\ref{eq1}) is the standard Feynman integration
contour; it will be deformed for a numerical evaluation as discussed below.

The vertex$+$reducible contribution is further divided into three parts: the zero-potential,
one-potential, and many-potential contributions,
\begin{align}\label{eq2}
\Delta g_{\rm vr} = \Delta g_{\rm vr}^{(0)} + \Delta g_{\rm vr}^{(1)} + \Delta g_{\rm vr}^{(2+)}\,.
\end{align}
This separation is induced by the following identity, which splits the integrand according to the
number of interactions with the binding Coulomb field in the electron propagators,
\begin{align}
 &\ G\,\delta V_{g}\, G \equiv \biggl[  G^{(0)}\,\delta V_{g}\, G^{(0)}\biggr]
 + \biggl[ G^{(0)}\,\delta V_{g}\, G^{(1)} + G^{(1)}\,\delta V_{g}\, G^{(0)} \biggr]
 \nonumber \\ &
+ \Biggl[ G\,\delta V_{g}\, G - G^{(0)}\,\delta V_{g}\, G^{(0)}
 - G^{(0)}\,\delta V_{g}\, G^{(1)} - G^{(1)}\,\delta V_{g}\, G^{(0)}\Biggr] \,,
\end{align}
where $G \equiv G(\vare) \equiv \sum_n |n\rbr\lbr n|/(\vare-\vare_n)$ is the bound-electron
propagator, $G^{(0)} \equiv G \bigr|_{Z = 0}$ is the free-electron propagator, and
$$
G^{(1)}(\vare) \equiv Z\, \left[ \frac{d}{dZ}\,G(\vare) \right]_{Z = 0}
$$
is the one-potential electron propagator.

In the present work, we will be concerned mainly with the numerical evaluation of $\Delta g_{\rm
vr}^{(2+)}$, since all other contributions were computed to the required accuracy in our previous
investigations \cite{yerokhin:04,yerokhin:10:sehfs}.

After performing integrations over the angular variables analytically as described in
Ref.~\cite{yerokhin:04}, we obtain the result that can be schematically represented as
\begin{align}
\Delta g_{\rm vr}^{(2+)} = \lim_{|\kappa_{\rm max}|\to\infty} \int_C d\omega\, \int_0^{\infty} dx\,dy\,dz\,
 \sum_{|\kappa| = 1}^{|\kappa_{\rm max}|} f_{|\kappa|}(\omega,x,y,z)\,,
\end{align}
where $x$, $y$, and $z$ are the radial integration variables, $|\kappa|$ is the absolute value of
the angular momentum-parity quantum number of one of the electron propagators, and $f_{|\kappa|}$
is the integrand. Summations over other angular quantum numbers are finite and absorbed into the
definition of $f_{|\kappa|}$.

The approach of the present work is to split $\Delta g_{\rm vr}^{(2+)}$ into two parts,
\begin{align}\label{eq3}
&  \Delta g_{\rm vr}^{(2+)} = \Delta g_{\rm vr,a}^{(2+)} + \Delta g_{\rm vr,b}^{(2+)}
 \nonumber \\ &
= \int_{C_{\rm LH,a}}\!\!\! d\omega\, \int_0^{\infty} dx\,dy\,dz\,
 \sum_{|\kappa| = 1}^{\kappa_{\rm a}} f_{|\kappa|}(\omega,x,y,z)
 \nonumber \\ &
+ \lim_{\kappa_{\rm max}\to\infty} \int_{C_{\rm LH,b}}\!\!\! d\omega\, \int_0^{\infty} dx\,dy\,dz\,
 \sum_{|\kappa| = \kappa_{\rm a}+1}^{\kappa_{\rm max}} f_{|\kappa|}(\omega,x,y,z)\,,
\end{align}
where $\kappa_a$ is an auxiliary parameter and $C_{\rm LH,a}$ and $C_{\rm LH,b}$ are two
integration contours used for the evaluation of the two parts of Eq.~(\ref{eq3}). In the present
work we used $\kappa_a=120$, which corresponds to the maximal value of $|\kappa|$ used in
Ref.~\cite{yerokhin:10:sehfs}, and $C_{\rm LH,a}$  being the same contour as used in that work. So,
the numerical evaluation of $\Delta g_{\rm vr,a}^{(2+)}$ was mostly analogous to the one reported
in Ref.~\cite{yerokhin:10:sehfs}, but we had to improve the accuracy of numerical integrations by
several orders of magnitude. In the updated numerical integrations, the extended Gauss-log
quadratures \cite{pachucki:14:cpc} were employed, alongside with the standard Gauss-Legendre
quadratures.

\begin{figure}
\centerline{
\resizebox{0.5\textwidth}{!}{%
  \includegraphics{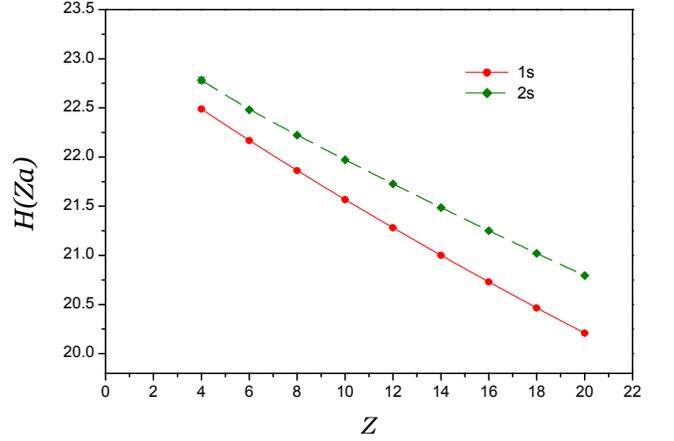}
}}
 \caption{The higher-order remainder function $H(\Za)$, defined by Eq. (\ref{eqh}), for the one-loop self-energy correction
 to the bound-electron $g$ factor of the $1s$ and $2s$ states.
 \label{fig:1}}
\end{figure}

We found it impossible to extend the partial-wave expansion significantly beyond the limit of
$\kappa_a=120$ within the same numerical scheme as used in Ref.~\cite{yerokhin:10:sehfs}. The
reason is that the integration contour $C_{\rm LH,a}$ used there, as well as in our previous works
\cite{yerokhin:02:prl,yerokhin:04}, involved computations of the Whittaker functions of the first
kind ${{M}_{\alpha,\beta}}(z)$ and their derivatives for large complex values of the argument $z$.
The algorithms we use \cite{yerokhin:99:pra} for computing ${{M}_{\alpha,\beta}}(z)$ become
unstable for large $\alpha$ (needed for large $\kappa$'s) and large and complex $z$, even when
using the quadruple-precision arithmetics. For this reason, in order to compute $\Delta g_{\rm
vr,b}^{(2+)}$, we had to switch to the contour $C_{\rm LH,b}$, which was originally introduced by
P.~J.~Mohr  in his calculations of the one-loop self-energy \cite{mohr:74:a,mohr:74:b}. The crucial
feature of this contour is that it involves the computation of the Whittaker functions
${{M}_{\alpha,\beta}}(z)$ and ${{W}_{\alpha,\beta}}(z)$ of the {\em real} arguments $z$ only. For
real arguments, the computational algorithms were shown \cite{mohr:74:b} to be stable even for very
large $\kappa$'s (and, hence, $\alpha$'s).

Specifically, the contours $C_{\rm LH,a}$ and $C_{\rm LH,b}$ consist of two parts, the low-energy
and the high-energy ones. The low-energy part extends along $(\Delta,0)$ on the lower bank of the
cut of the photon propagator of the complex $\omega$ plane and along $(0,\Delta)$ on the upper bank
of the cut. The high-energy part consists of the interval $(\Delta,\Delta+i\infty)$ in the upper
half-plane and the interval $(\Delta,\Delta-i\infty)$ in the lower half-plane. The difference
between $C_{\rm LH,a}$ and $C_{\rm LH,b}$ is only in the choice of the parameter $\Delta$. For
$C_{\rm LH,a}$, we use $\Delta = \Za\,\vare_a$ (the same choice as in our previous works
\cite{yerokhin:02:prl,yerokhin:04,yerokhin:10:sehfs}), whereas for $C_{\rm LH,b}$, we use $\Delta =
\vare_a$ (the Mohr's choice). Detailed discussion of the integration contour and the analytical
properties of the integrand can be found in the original work~\cite{mohr:74:a}.

We found that the price to pay for using the contour $C_{\rm LH,b}$ was the oscillatory behavior of
the integrand as a function of the radial variables for $\omega \sim \vare_a$. Because of this, we
had to employ very dense radial grids for numerical integrations, which made computations rather
time-consuming.

The largest error of the numerical evaluation of Eq.~(\ref{eq3}) comes from the termination of the
infinite summation over $|\kappa|$ and the estimation of the tail of the expansion. In the present
work, we performed the summation over $|\kappa|$ {\em before} all integrations and stored the
complete sequence of partial sums, to be used for the extrapolation performed on the last step of
the calculation. The convergence of the expansion was monitored; in the cases when the series
converged to the prescribed accuracy (i.e., the relative contribution of several consecutive
expansion terms was smaller than, typically, $10^{-11}$ for $\Delta g_{\rm vr,a}^{(2+)}$ and
$10^{-6}$ for $\Delta g_{\rm vr,b}^{(2+)}$), the summation was terminated. This approach reduced
the computation time considerably as compared to our previous scheme \cite{yerokhin:04}, where the
summation over $|\kappa|$ was performed {\em after} all integrations. If the convergence of the
partial-wave expansion had not been reached, the summation was extended up to the upper cutoff
$\kappa_{\rm max} = 450$.

The remaining tail of the series was estimated by analyzing the $|\kappa|$-dependence of the
partial-wave expansion terms {\em after} all integrations. We fitted the last $m$ expansion terms
(typically, $m = 20$) to the polynomial in $1/|\kappa|$ with 1-3 fitting parameters,
$$\delta S_{|\kappa|} = c_0/|\kappa|^3 + c_1/|\kappa|^4+\ldots\,.$$
The uncertainty of the extrapolation was estimated by varying the cutoff parameter $\kappa_{\rm
max}$ by 20\% and multiplying the resulting difference by a conservative factor of $1.5$. This
procedure usually led to the expansion tail estimated with an accuracy of about 10\%.

We observed an interesting feature, namely, that the tail of the expansion, with a high accuracy,
is the same for the $1s$ and for the $2s$ states. E.g., for $Z = 4$, we find the expansion tail of
$\delta g(1s) = -1.88\,(19)\times 10^{-12}$ and $\delta g(2s) = -1.88\,(19)\times 10^{-12}$; for $Z
= 16$, we obtain $\delta g(1s) = -3.00\,(27) \times 10^{-11}$ and $\delta g(2s) = -3.01\,(27)
\times 10^{-11}$. We do not know the reason for this but such an agreement shows a high degree of
consistency of our numerical calculations for the $1s$ and $2s$ states.

Our numerical results for the self-energy correction to the bound-electron $g$ factor of the $1s$
and $2s$ states of hydrogen-like ions are presented in Table~\ref{tab:results}. The values for the
irreducible part $\Delta g_{\rm ir}$ are taken from our previous investigations (from
Ref.~\cite{yerokhin:10:sehfs} for $Z \le 12$ and from Ref.~\cite{yerokhin:04} otherwise). Using
results of Ref.~\cite{yerokhin:04}, we introduced small corrections that accounted for a different
value of the fine-structure constant used in that work. In Table~\ref{tab:results} we also present
values of the higher-order remainder function $H(\Za)$, obtained after separating out all known
terms of the $\Za$ expansion \cite{pachucki:04:prl,pachucki:05:gfact} from our numerical results,
\begin{align}\label{eqh}
\Delta g_{\rm SE} = &\ \frac{\alpha}{\pi}\Biggl[ 1 + \frac{(\Za)^2}{6n^2} + \frac{(\Za)^4}{n^3}\,
  \biggl\{ \frac{32}{9}\,\ln[(\Za)^{-2}]+ b_{40}\biggr\}
 \nonumber \\ &
  + \frac{(\Za)^5}{n^3}\,H(\Za)\Biggr]\,,
\end{align}
where $b_{40}(1s) = -10.236\,524\,32$ and $b_{40}(2s) = -10.707\,715\,60$.  The results for the
higher-order remainder function are plotted in Fig.~\ref{fig:1}.

Our calculation represents an improvement in accuracy over previous works by about two orders of
magnitude. Table~\ref{tab:comp} shows the comparison of various calculations for carbon. It is
gratifying to find that all results are consistent with each other within the given error bars.

In the present work, we performed direct numerical calculations for ions with $Z\geq 4$. For
smaller $Z$, numerical cancelations in determining the higher-order remainder become too large to
make numerical calculations meaningful. Instead of direct calculations, we extrapolated the
numerical values presented in Table~\ref{tab:results} for $H(\Za)$ down towards $Z\to 0$. Doing
this, we assumed the following ansatz for $H(\Za)$, which was inspired by the expansion of the
one-loop self-energy for the Lamb shift,
\begin{align} \label{eqza}
H(\Za) \approx &\  c_{00} + (\Za)\biggl\{\ln^2[(\Za)^{-2}]\,c_{12}
 \nonumber \\ &
 + \ln[(\Za)^{-2}]\,c_{11}
+c_{10}\biggr\} + (\Za)^2\,c_{20}\,.
\end{align}
For the $2s\,$-$1s$ difference, we use the form (\ref{eqza}) with $c_{12} = 0$, assuming the
leading logarithm to be state-independent. The extrapolated results are presented in
Table~\ref{tab:extr}. The uncertainties quoted for our fitting results are obtained under the
assumption that the logarithmic terms in the next-to-leading order of the $\Za$ expansion of
$H(\Za)$ comply with Eq.~(\ref{eqza}). If we introduce, e.g., a cubed logarithmic term into
Eq.~(\ref{eqza}), our estimates of uncertainties would increase by about a factor of 2.

In summary, we reported calculations of the one-loop self-energy correction to the bound-electron
$g$ factor of the $1s$ and $2s$ state of light hydrogen-like ions, performed to all orders in the
binding nuclear strength parameter $\Za$. The relative accuracy of the results obtained varies from
$1\times 10^{-10}$ for $Z = 4$ to $3\times 10^{-9}$ for $Z = 20$. Our results agree well with the
previously published values but their accuracy is by about two orders of magnitude higher.

\section*{Acknowledgement}

V.A.Y. acknowledges support by the Ministry of Education and Science of the Russian Federation
Grant No. 3.5397.2017/BY.

\begin{table*}
\caption{One-loop self-energy correction to the bound-electron $g$ factor for the $1s$ and $2s$ states of H-like ions,
multiplied by $10^6$. The value of the fine-structure constant used in the calculation is $\alpha^{-1} = 137.036$. \label{tab:results}}
\begin{ruledtabular}
\begin{tabular}{l......}
 \multicolumn{1}{c}{$Z$}   & \multicolumn{1}{c}{$\delta g_{\rm ir}$}  & \multicolumn{1}{c}{$\delta g_{\rm vr}^{(0)}$}  & \multicolumn{1}{c}{$\delta g_{\rm vr}^{(1)}$}  & \multicolumn{1}{c}{$\delta g_{\rm vr}^{(2+)}$}  & \multicolumn{1}{c}{Total}  & \multicolumn{1}{c}{$H(\Za)$} \\
\hline\\[-5pt]
 \multicolumn{1}{c}{$1s$}\\[2pt]
  4 & 17.216\,x132\,6 & 2\,300.99x7\,357\,2 & 4.459\,x445\,1 & 0.502\,5x85\,3\,(2)  & 2\,323.175\,x520\,1\,(2) & 22.48x7\,(4) \\
  6 & 34.064\,x668\,1 & 2\,280.73x7\,822\,5 & 7.795\,x347\,6 & 1.074\,5x84\,5\,(4)  & 2\,323.672\,x422\,7\,(4) & 22.16x6\,(1) \\
  8 & 54.781\,x703\,2 & 2\,256.69x7\,710\,8 & 11.165\,x707\,2 & 1.797\,0x02\,1\,(7) & 2\,324.442\,x123\,2\,(7) & 21.86x1\,4\,(5) \\
 10 & 78.743\,x788\,6 & 2\,229.82x6\,130\,5 & 14.349\,x045\,7 & 2.617\,5x42\,(1)    & 2\,325.536\,x507\,(1) & 21.56x6\,3\,(2) \\
 12 & 105.511\,x685\,3 & 2\,200.79x8\,139\,6 & 17.216\,x228\,5 & 3.483\,7x62\,(2)   & 2\,327.009\,x815\,(2) & 21.27x9\,5\,(1) \\
 14 & 134.760\,x370\,(3) & 2\,170.11x9\,540\,6 & 19.693\,x048\,7 & 4.344\,5x33\,(2) & 2\,328.917\,x492\,(4) & 21.00x0\,5\,(1) \\
 16 & 166.242\,x092\,(3) & 2\,138.18x2\,197\,1 & 21.740\,x396\,0 & 5.150\,8x96\,(3) & 2\,331.315\,x581\,(4) & 20.72x9\,03\,(8) \\
 18 & 199.764\,x465\,(3) & 2\,105.29x6\,890\,3 & 23.342\,x539\,0 & 5.856\,5x57\,(3) & 2\,334.260\,x452\,(5) & 20.46x5\,04\,(5) \\
 20 & 235.176\,x430\,(4) & 2\,071.71x4\,411\,8 & 24.499\,x708\,2 & 6.418\,1x72\,(4) & 2\,337.808\,x723\,(6) & 20.20x8\,29\,(4) \\[2pt]
\hline\\[-5pt]
 \multicolumn{1}{c}{$2s$}\\   [2pt]
  4 &  5.053\,x860\,6 & 2\,315.98x8\,700\,0 & 1.337\,x877\,4 & 0.524\,6x54\,7\,(3) & 2\,322.905\,x092\,6\,(3) & 22.78x\,(4) \\
  6 & 10.186\,x275\,6 & 2\,309.23x0\,319\,4 & 2.428\,x638\,8 & 1.173\,1x19\,8\,(4) & 2\,323.018\,x353\,7\,(4) & 22.48x\,(1) \\
  8 & 16.629\,x185\,2 & 2\,300.88x5\,484\,0 & 3.600\,x384\,8 & 2.070\,6x36\,5\,(7) & 2\,323.185\,x690\,5\,(7) & 22.22x1\,(4) \\
 10 & 24.208\,x822\,9 & 2\,291.21x6\,226\,1 & 4.777\,x844\,0 & 3.210\,2x48\,(1)    & 2\,323.413\,x141\,(1) & 21.97x2\,(2) \\
 12 & 32.796\,x260\,4 & 2\,280.41x6\,694\,2 & 5.909\,x016\,2 & 4.584\,9x79\,(2)    & 2\,323.706\,x950\,(2) & 21.72x7\,(1) \\
 14 & 42.290\,x335\,(2) & 2\,268.63x8\,681\,0 & 6.956\,x360\,9 & 6.188\,0x74\,(2)  & 2\,324.073\,x450\,(3) & 21.48x6\,(1) \\
 16 & 52.608\,x698\,(2) & 2\,256.00x5\,047\,5 & 7.892\,x134\,5 & 8.013\,1x44\,(3)  & 2\,324.519\,x024\,(4) & 21.25x0\,5\,(6) \\
 18 & 63.682\,x502\,(3) & 2\,242.61x7\,687\,9 & 8.695\,x612\,8 & 10.054\,2x80\,(4) & 2\,325.050\,x083\,(5) & 21.02x0\,3\,(4) \\
 20 & 75.453\,x014\,(4) & 2\,228.56x2\,653\,6 & 9.351\,x303\,3 & 12.306\,1x20\,(5) & 2\,325.673\,x091\,(6) & 20.79x4\,9\,(3) \\
\end{tabular}
\end{ruledtabular}
\end{table*}

\begin{table}
\caption{The higher-order remainder $H(\Za)$ for the $1s$ and $2s$ states of H-like carbon ($Z = 6$), in different calculations. \label{tab:comp}}
\begin{ruledtabular}
\begin{tabular}{lll}
\multicolumn{1}{c}{$H_{1s}(6\alpha)$}   & \multicolumn{1}{c}{$H_{2s}(6\alpha)$} & Ref.  \\ \hline\\[-5pt]
22.166\,(1) & 22.48\,(1)  & This work \\
22.18\,(9)  & 22.5\,(1.3) & \cite{yerokhin:10:sehfs} \\
22.16\,(1)  &              & \cite{pachucki:04:prl}$^a$ \\
22.2\,(2)   & 18.(13.)    & \cite{yerokhin:02:prl,yerokhin:04} \\
22.\,(2.)   &              & \cite{beier:00:pra}\\
\end{tabular}
$^a$ extrapolation of the numerical data from \cite{yerokhin:02:prl}.
\end{ruledtabular}
\end{table}

\begin{table}
\caption{Extrapolated values of the higher-order remainder $H(\Za)$ for small $Z$. \label{tab:extr}}
\begin{ruledtabular}
\begin{tabular}{lll}
 \multicolumn{1}{c}{$Z$}   & \multicolumn{1}{c}{$H_{1s}$}  & \multicolumn{1}{c}{$H_{2s}-H_{1s}$}\\
\hline\\[-5pt]
0 & 23.6\,(5)  & 0.12\,(5) \\
1 & 23.08\,(9) & 0.16\,(3) \\
2 & 22.85\,(3) & 0.20\,(2) \\
\end{tabular}
\end{ruledtabular}
\end{table}


\end{document}